\documentclass[12pt,a4paper,reqno]{article}
\usepackage[english]{babel}
\usepackage[utf8]{inputenc} 
\usepackage{amssymb,amsmath}
\usepackage{fontenc,indentfirst, delarray, amsfonts,amssymb}
\usepackage{graphicx,graphics,epstopdf,epsf}
\DeclareGraphicsExtensions{eps,ps,p.gz,pdf}
\usepackage[pdfstartview=FitH]{hyperref}

\makeatletter
\renewcommand{\section}{\@startsection {section}{1}{\z@}%
             {-3.5ex \@plus -1ex \@minus -.2ex}%
             {2.3ex \@plus.2ex}%
             {\normalfont\normalsize\sffamily\bfseries}}
\renewcommand{\subsection}{\@startsection {subsection}{1}{\z@}%
             {-3.5ex \@plus -1ex \@minus -.2ex}%
             {2.3ex \@plus.2ex}%
             {\normalfont\normalsize\sffamily\emph}}

\numberwithin{equation}{section}

\newcommand{\norm}[1]{\left\lVert#1\right\rVert}
\newcommand{\abs}[1]{\lvert#1\rvert}

\def\lp{\left(} 
\def\rp{\right)} 
\def\dm{\lp\begin{array}}	
\def\fm{\end{array}\rp}
\def\m2{M_2 \lp \cc \rp}

\def\cc{{\mathbb{C}}}	
\def\rr{{\mathbb{R}}}

\def\aa{{\mathcal A}}

\def\oo{{\mathcal O}}

\begin{document}
\title{Conformal mapping of Unruh temperature}
\author{Pierre Martinetti{\footnote{martinetti@physik.uni-goe.de}} 
\\
\small{Institut f\"ur Theoretische Physik, Georg-August Universit\"at,
 G\"ottingen,}\\
\small{Dipartimento di fisica-GTC, Universit\`a di Roma "La Sapienza"}}
\small{\maketitle}

\begin{abstract}
Thanks to a local interpetation of the KMS condition, the mapping from (unbounded)  wedge regions of Minkowski space-time to (bounded)
  double-cone regions is extended to the Unruh temperature associated to relevant observers in both regions.
A previous result,  the diamond's temperature, is shown to be proportional to the inverse of the conformal factor (Weyl rescaling of the metric) of this map. One thus explains from a mathematical point of view why an observer with finite lifetime experiences the vacuum as a thermal state whatever his acceleration, even vanishing.

\end{abstract}



\section{Introduction}

 Consider a quantum field theory on Minkowski space-time
$M$ with a vacuum state $\Omega$. For an inertial observer with {\it
  infinite lifetime}, i.e. whose
line of universe in Cartesian coordinates  $x^\mu = (t, \overrightarrow{x})$ is 
$
(t\in\rr, \overrightarrow{x}= \text{constant}),
$
 the vacuum is a zero particle state. However for an observer with
a constant acceleration $a$ and proper time  $\tau$,
\begin{equation}
\label{paramprop}
x^\mu(\tau) = \lp a^{-1} \text{sh} a\tau,  a^{-1} \text{ch} a\tau, 0
,0\rp\quad\quad \tau\in \left]-\infty,+\infty\right[ ,
\end{equation}
the same vacuum is seen 
as a thermal bath of quanta
of the fields at a temperature $T_U = \frac{a}{2\pi}$.
 Known as Unruh temperature, $T_U$ can be derived in various ways \cite{unruh,davies,fulling,dewitt} including an 
approach based on
the algebra of local observables accessible for the uniformly accelerated observer  
  \cite{bisognano1,bisognano2,sewell1,sewell2}. 

It is quite natural to wonder if there exists other classes of observers, i.e. other trajectories than (\ref{paramprop}), for which the vacuum appears as a thermal state. This question can be addressed in various ways. Most often \cite{louko} the authors consider a quantum system interacting with the vacuum along a chosen trajectory. The point is then to determine whether the system gets excited
 as if it were at rest in a thermal bath of quantas. The result heavily depends 
on the chosen trajectory as well as on the form of the vacuum interaction.  In \cite{diamonds}, inspired by the \emph{thermal time hypothesis} of Connes and Rovelli  \cite{carloconnes}, we addressed the question in a different way, following the ``local algebra point of view'' on Unruh effect. 
Under the assumption that the field
is  conformal invariant we found that the vacuum is still thermal at a temperature $T_D$ for a uniformly accelerated observer
with \emph{finite} lifetime. However, in contrast with the eternal case, $T_D$ 
does not vanishes for an inertial observer.

\begin{figure}[bt]
\begin{center}
\mbox{\rotatebox{0}{\scalebox{.7}{\includegraphics{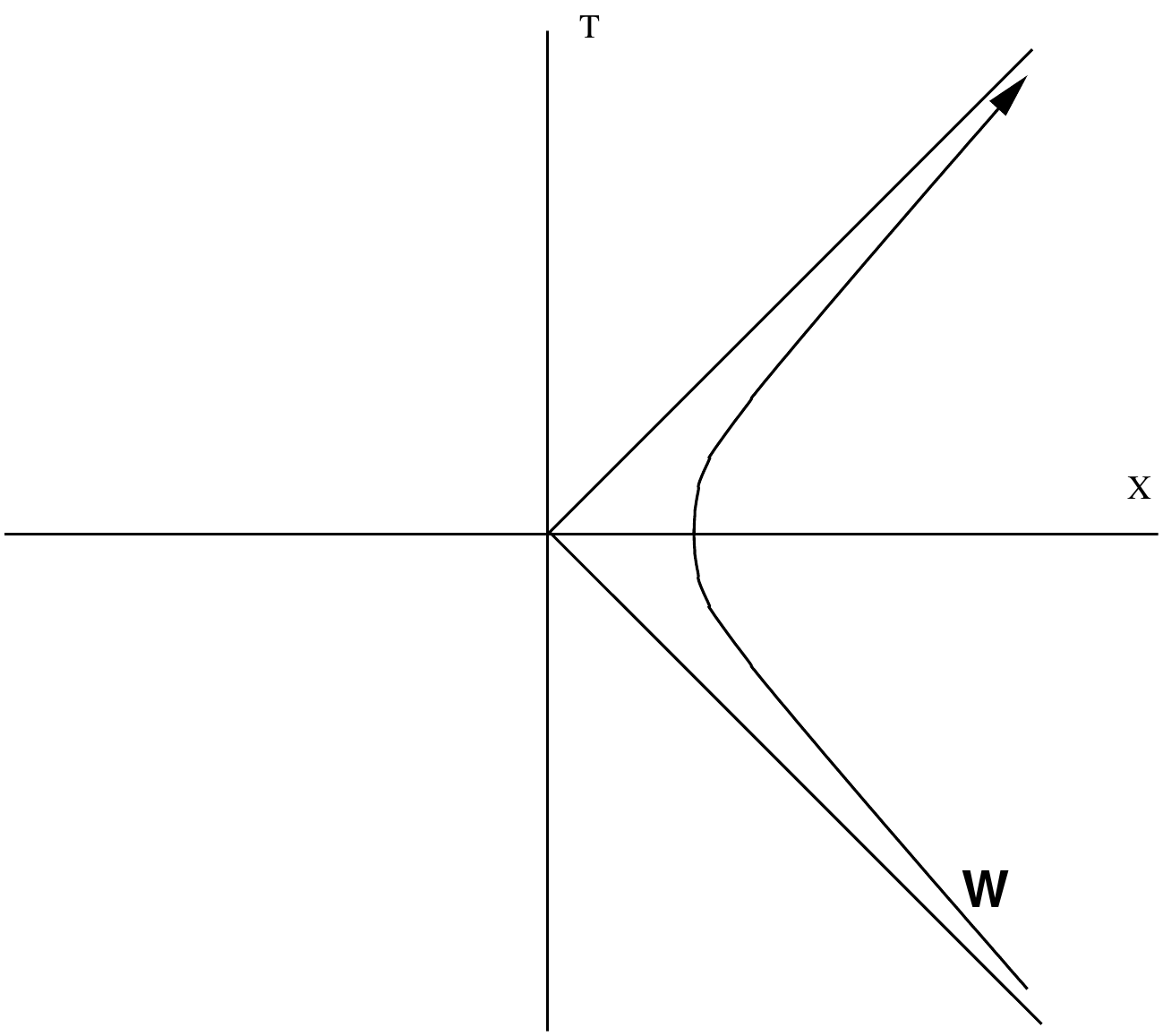}}}}
\caption{\small{The Rindler wedge $W$ and the trajectory $H_W$ of a uniformly accelerated observer.}}
\label{figwedge}
\end{center}
\end{figure}

\begin{figure}[htbt]
\begin{center}
\mbox{\rotatebox{0}{\scalebox{.7}{\includegraphics{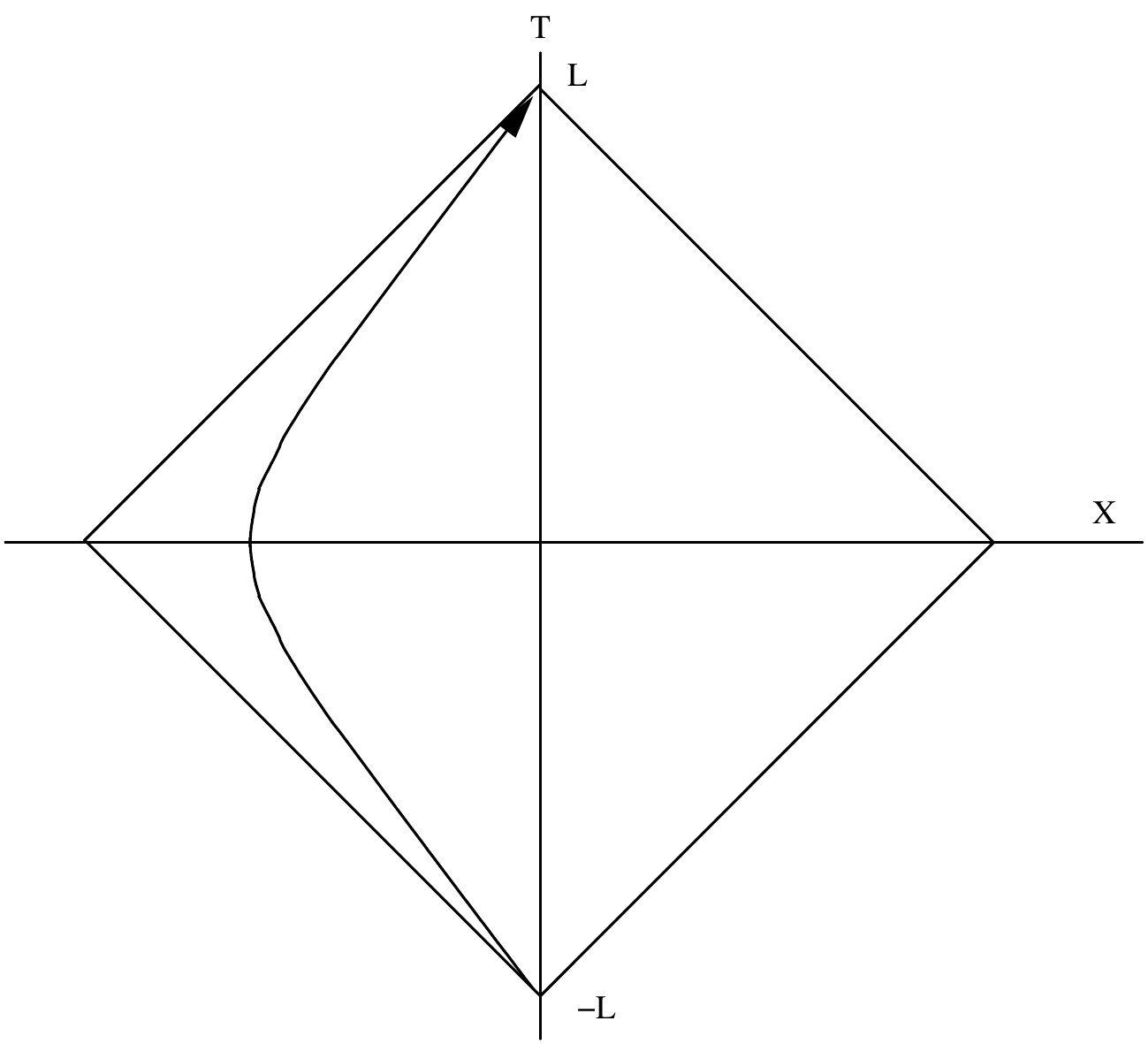}}}}
\caption{\small{A double-cone, or diamond, region $D$ of size $L$ together
  with the trajectory $H_D$ of a uniformly accelerated observer with finite lifetime.}}
\label{figdiam}
\end{center}
\end{figure}

The present note aims at re-deriving this result on some geometrical basis, providing a mathematical
explanation to the non-vanishing of $T_D$ for zero acceleration. Explicitly $T_D$ is shown to be proportional to the inverse of the conformal factor $C$
of the map $\varphi$ that sends a wedge region $W$ of Minkowski space-time $M$,
\begin{equation}
W = \left\{ \abs{\overrightarrow{x}} > \abs{t}\right\},\label{eq:3}
\end{equation} 
to a double-cone region $D$,
\begin{equation}
\label{eqdcone}
\abs{\overrightarrow{x}} + \abs{t} < L
\end{equation}
where $L$ is a real constant (see figures \ref{figwedge}, \ref{figdiam}).
The non-vanishing of $T_D$ for zero acceleration then reflects that $C$ is everywhere finite, which is natural regarding a map sending an unbounded region of Minkowski space-time
to a bounded one.

Section 2 is a review on Unruh effect within the local algebraic quantum field theory framework and recalls why 
wedges region are relevant. Section 3 
studies the adaptation to non-eternal uniformly accelerated observers, for which double-cones are the relevant regions, and establishes the link between $T_D$ and
the conformal factor $C$. Last section contains discussion and interpretation.
 \\

Note that hereafter we do not make any distinction between observer
and thermometer. We consider the temperature defined at a point,
although from an experimental point of view one should also take into
account the spatial extension of the measuring device. In curved
space-time in particular this point may be delicate, as illustrated in
 \cite{buchholz}, since the
temperature 
can vary in a neighborhood of the geodesic followed by the observer,
eventually not allowing for a consistent thermal interpretation. We do
not take this point into consideration here.
\\

We work in units $c=\hbar = 1$ and with signature $+---$.

\section{\label{Unruh effect, the algebraic way} Unruh effect, the algebraic way}

Let us recall that in algebraic quantum field theory \cite{haag}, as soon as an open region $\oo$ of Minkowski space-time has
a non-void causal complement, the associated algebra of local observables $\aa(\oo)$ comes equipped with a canonical 1-parameter group of
automorphisms associated to $\Omega$ (the modular
group \cite
{connesmodul,tomtak}), 
\begin{equation}
s\in\rr \mapsto \sigma(\oo)_s\in \text{Aut}(\aa(\oo)).
\end{equation}
Furthermore an important result of modular theory states that $\Omega$ satisfies with respect to $\sigma(\oo)$ the
same properties that does a thermal equilibrium state at temperature $-1$ with respect to time evolution, namely the KMS
condition
\begin{equation}
  \label{KMS}
 \omega ((\sigma_s A) B) = \omega(B (\sigma_{s-i}A)) \quad\quad \forall A,B\in\aa(\oo)
\end{equation}
where $\omega(.)$ stands for $\langle \Omega, . \Omega\rangle$ and we omit $\oo$ to lighten notation.
In an abstract sense \eqref{KMS} indicates that for an observer whose time-flow is given by $\sigma(\oo)$ the vacuum $\Omega$ has all the properties
of an equilibrium state at temperature $-1$. To make this statement physically meaningful one has to determine whether $\sigma(\oo)$  actually corresponds to a concrete physical situation. This is not an easy task since there is a priori no reason that
the automorphism $\sigma(\oo)_s$ can be written as $a\mapsto e^{-iHs}ae^{iHs}$ for some physically meaningful Hamiltonian $H$. Nevertheless for some open regions $\oo$ the identification of $\sigma(\oo)$ as a physical flow of time is possible: by definition the algebra of local observable $\aa(\oo)$ 
always carries a representation of the Poincare group, and it turns out that for some $\oo$, e.g wedges or some double-cones, 
$\sigma(\oo)$ is 
precisely generated by a Poincare transformation. Consequently the modular group acquires a geometrical action on space-time itself and
 \eqref{KMS} can be interpreted as follows: the vacuum $\Omega$ is a thermal state at temperature $-1$ for an observer
whose line of universe coincides with an orbit $H_{\oo}$ of the geometrical action of $\sigma_{\oo}$ on $M$. 
However in order to identify $s$ to the proper time $\tau$ of some observer one has to check that $\partial_s$ behaves as a well-defined $4$-velocity, namely that it has norm $1$. Therefore it is convenient to introduce the normalized flow
\begin{equation}
\label{ddtdds}
  \partial_t \doteq - \beta^{-1}\partial_s
\end{equation}
where
\begin{equation}
  \label{defbeta}
  \beta \doteq \norm{\partial_s}.
\end{equation}
 Writing $\alpha_{-\beta s}\doteq \sigma_s$, (\ref{KMS}) then yields
\begin{equation}
\label{kmstot}\omega ((\alpha_{-\beta s} A) B) = \omega(B (\alpha_{-\beta (s-i)}A)) = \omega(B (\alpha_{-\beta s +i\beta }A))
\end{equation}
which is precisely the KMS condition characterizing a thermal state at temperature $\beta^{-1}$ for a system with time parameter
\begin{equation}
  \label{dtparam}
  t \doteq -\beta s.
\end{equation}
To summarize when $\sigma(\oo)$ has a geometrical action the vacuum $\Omega$ is a thermal state at temperature $\beta^{-1}$ for an observer with proper time $-\beta s$.  The temperature appears as the normalization factor\footnote{One might  note an analogy with Tolman's law
in cosmology that relates the temperature of a body seen from two two observers with different redshift. However here there exists no observer who sees the vacuum at temperature $-1$. The physically meaningful flow is $\partial_t$, not $\partial s$.} required to make the abstract modular flow $\partial_s$
 a physical flow $\partial_t$. 

For example when $\oo = W$  the modular group $\sigma(W)$ 
is found to be generated by boosts \cite{bisognano1,bisognano2} so that the orbit of the point $(0,a^{-1},0,0)\in M, a>0$ is an hyperbola $H_W$,
\begin{equation}
\label{paramodul}
x^\mu(s) = \lp -a^{-1} \text{sh}\, 2\pi s,  a^{-1} \text{ch}
\,2\pi s, 0
,0\rp,\quad s\in \left] -\infty, + \infty\right[ .
\end{equation}
One easily computes
\begin{equation}
\label{unruhtemp}
\beta = \frac{2\pi}{a} = T_U^{-1}
\end{equation}
which is constant on each $H_W$. Noticing that $H_W$ (parametrized by $s$) coincides with the line of universe \eqref{paramprop} (parametrized by $\tau$)
up to the identification $a\tau = -2\pi s$, one observes that the thermal time $t$ defined in \eqref{dtparam} is nothing but $\tau$. In other words the vacuum is KMS with respect to the time evolution of the uniformly accelerated observer, hence
Unruh effect.

%
%
%

\section{From wedges to double-cones}

$W$ is relevant in the derivation of Unruh temperature for it represents the whole and only region of $M$
with whom an eternal uniformly accelerated observer can
interact, that is he cannot exchange information with somebody
 outside $W$. In this sense the edge of the wedge acts like
an horizon, and there is indeed an analogy between Unruh temperature
and Hawking radiative temperature $T_H$ of black holes: for eternal
black holes $T_H$ identifies to $T_U$ for an observer
moving on the horizon, with acceleration the surface gravity
of the black hole. 
%
On the contrary for an inertial observer 
the edges of the wedge are sent to
infinity,  $W$ tends to $M$, there is no longer horizon and $T_U$ vanishes.
The situation is different for a non-eternal observer. There the
existence of an horizon is no longer linked to the non vanishing of
the acceleration. Whether he is accelerated or inertial, a non-eternal
observer experienced a causal horizon given by the intersection of the future cone of his birth with the
past cone of his death. Up to a Lorenz transformation  this ``life
horizon'' \cite{thermalhorizon} can always be taken as the boundary of a
double-cone (or diamond region) $D$ whose equation is given in \eqref{eqdcone}.
If one believes, as indicated by the analogy with
Hawking temperature, that the thermalization of the vacuum is due to
the presence of an horizon, there should be no reason for the life
horizon not to be thermal. 

A naive guess to define a vacuum-temperature for a non-eternal observer with acceleration $a'$ would be to take Unruh formula and define the temperature as 
$\frac{a'}{2\pi}$. However the proportionality between the temperature and the acceleration comes from the modular structure of $W$.
 There is no reason the same proportionality holds for double-cones. The only thing we know for double-cones is that the vacuum is KMS with respect 
to the double-cone modular group $\sigma(D)$. To turn this mathematical observation into a physical result, one should adapt the strategy used to derive Unruh temperature, namely 1) check if $\sigma(D)$ has a geometrical action on $M$, and whether the corresponding orbits $H_D$ coincides with trajectories of physical observers. 2) 
identify the thermal time defined in (\ref{dtparam}) to the proper time $\tau'$ of this observer so that to obtain, via the KMS condition,
that the vacuum is thermal with inverse temperature 
\begin{equation}
  \label{ratio}
  \beta = \norm{\partial_{s'}} = \abs{\frac{d\tau'}{d s'}}
\end{equation}
where, to avoid confusion, we write $s'$ the modular parameter associated to $\sigma(D)$. 
Without any computation, it is obvious that this temperature cannot be proportional to a constant acceleration, for  an observer whose line of universe $H_D$ entirely
lies within a double-cone has a proper
time that ranges within a finite interval $\left]-\tau'_0, \tau'_0\right[$ whereas the
modular parameter still runs all over $\rr$. So the ratio (\ref{ratio}) that
measures the difference between two parametrizations of $H_D$ can no longer be a 
constant, but should rather be a function of $\tau'$.

\subsection{Modular flow in double-cones}
\label{wdc}

\begin{figure}[ht]
\begin{center}
\mbox{\rotatebox{0}{\scalebox{1}{\includegraphics{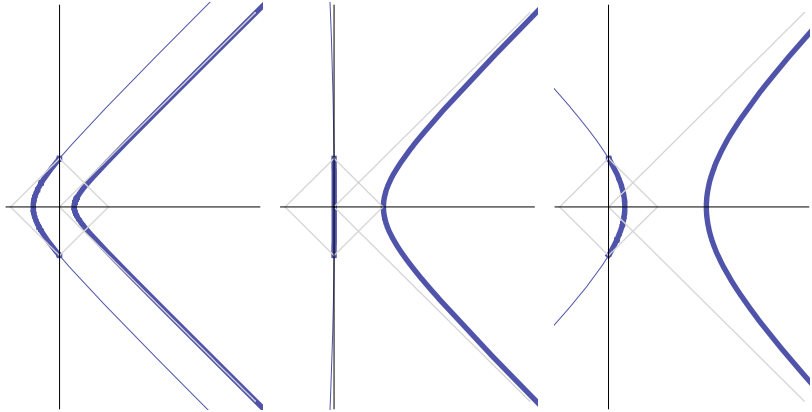}}}}
\caption{Hyperbolas in $W$ ands their image  in $D$ (thick lines) under the conformal map $\varphi$. 
Note that the image in $D$ is part of a hyperbola (thin line).}
\label{wdfig}
\end{center}
\end{figure}

In explicitly computing $\beta$ for the double-cone  the hard part is to
determine the modular group $\sigma(D)$. Fortunately this has been done time ago by Hislop and
Longo \cite{hislop}. Before stating their result, let us recall that their starting point is the observation
that double-cones come from wedges by a conformal mapping. Explicitly the map from $W$ to the double-cone of
size $1$ is the composition
\begin{equation}
\label{wdconfo}
S_1 \circ T_{-1} \circ \rho \circ T_\frac{1}{2} \circ \Lambda_\frac{1}{2}
\end{equation}
where $\Lambda_{\frac 12}$ is the dilation $x^\mu\mapsto \frac 12 x^\mu$,
$T_\lambda$ is the translation $x^1 \mapsto x^1 + \lambda$, $\rho$ is
the relativistic inversion
$$x^\mu\mapsto \frac{-x^\mu}{\abs{x}^2}$$ with 
$\abs{x}^2 = t^2 - \abs{\overrightarrow{x}}^2$ 
the norm of the $4$-vector position and $S_1$ is the symmetry $x^1 \mapsto -x^1$. 
Composing
(\ref{wdconfo}) on the left with a dilation of amplitude $L$ one
defines a conformal transformation $\varphi$ that maps $x\in W$ with 
coordinates $x^\mu = (t,x^i)$ to  ${x'}^\mu = (t',{x'}^i)$ in the double-cone
$D$ of size $L$. Explicitly 
\begin{equation}
{x'}^{\mu} =\frac{L}{N(x)}\lp \begin{array}{c}
2 t\\
-1 - \abs{x}^2\\
2x^2\\
2x^3
\end{array}\right)
\end{equation}
with{\footnote{It is immediate to check that $x'$ belongs to $D$ since, noting that $N$ is positive on $W$,
the condition $
\abs{{x'}^0} + \abs{\overrightarrow{x'}} \leq L\,
$
 is equivalent to
$
(x^1 - \abs{t}) N(x)> 0,
$
which is true for any $x$ in $W$.}}
\begin{equation}
  \label{eq:1}
  N(x)\doteq 1+ 2x^1 -\abs{x}^2.
\end{equation}
In particular the unbounded accelerated trajectory (\ref{paramprop}) is mapped to  
\begin{equation}
  \label{eq:5}
 t' =\frac{2La^{-1}\sinh a\tau}{N(x)}, \,  x'^1 =\frac L{N(x)} (-1+a^{-2}) 
\end{equation}
with
\begin{equation}
N(x) = 1 + 2a^{-1} \cosh a\tau + a^{-2}.\label{eq:6}
\end{equation}
 Obviously when $\tau\rightarrow \pm \infty$ then $t'\rightarrow \pm L$ and
$x'^1$ vanishes. Moreover a little bit of algebra shows that
 \begin{equation}
\label{hyperbprim}
t'^2 -(x'^1 -K)^2 = -a'^{-2}
\end{equation}
where
\begin{equation}
  \label{eq:2}
  a' =  \frac{-1 + a^2}{2 a L}
\end{equation}
and $K= -a'^{-1}\sqrt{1+a'^2 L^2}$.
Thus (\ref{paramprop}) is mapped to a branch of hyperbola whose closure contains the top and bottom of $D$. To say it differently, through $\varphi$ an eternal uniformly accelerated observer with acceleration $a$
is mapped to a uniformly accelerated observer with finite lifetime and acceleration $a'$. Note that when $a$ runs from $+\infty$ to $0$, then $a'$ runs from $+\infty$ to $-\infty$, see figure \ref{wdfig}. In particular the hyperbola in the wedge with acceleration $a=1$ maps to the inertial trajectory in the double-cone.
 
Having noticed that wedges and double-cones are conformaly equivalent, one can map the modular theory from $W$ to $D$ as soon as the fields
satisfy some conformal symmetries compatible with the algebraic structure of local observables. These conditions are rather strong (see \cite{hislop} for details)
and are verified for instance for massless  free fields. For massive fields the modular group $\sigma(D)$ is still an open (and difficult) question. It is expected that its action is not purely geometric \cite{saffary,saffary2}. In our case, assuming the field has the required conformal invariance, it turns
 out that 
\begin{equation}
\label{sigmad}
\sigma(D) = \varphi \circ \sigma(W) \circ\varphi^{-1}.
\end{equation}
The orbit under the action of $\sigma(W)$ of $\varphi^{-1}(x')$ for  some $x'\in D$ is an hyperbola $H_W$ corresponding to an acceleration $a$. The orbit $H_D$ of $x'$ under the action of $\sigma(D)$ is thus $\varphi(H_W)$ which, as explained in the precedent section, coincides with the trajectory of a
 non-eternal uniformly accelerated observer (\ref{hyperbprim}) with acceleration $a'$.
Specifically for $x'=(0,u,0,0)$ with $u\in ]-L,L[$, (\ref{hyperbprim}) yields $\abs{u-K} = \abs{a'^{-1}}$ so that
\begin{equation}
   \label{eq:4}
   a' = \pm\frac{2u}{L^2-u^2}
 \end{equation}
 where the sign is opposite to the one of $u${\footnote{A uniformly accelerated observer whose trajectory passes through the extremities of the double-cone
and at the point $(0,u>0,0,0)$ as an acceleration oriented towards the negative $x$-axis.}}.
 Having shown that $H_D$ coincides with a physical trajectory, it makes sense to apply (\ref{ratio}) to define the temperature of the vacuum.

 \subsection{The conformal factor as a temperature}
 The vector $\partial_{s'}$ tangent to $H_D$ is the push-forward by the conformal  map $\varphi$ of the vector tangent to
$H_W$, $\partial_{s'} =  \varphi_* \partial_{s}.$
Therefore
\begin{equation}
\label{betaprim}
\beta(x') \doteq \norm{\varphi_* {\partial_{s}}_{\lvert_{x'}}} =
\sqrt{g(\varphi_* {\partial_{s}}_{\lvert_{x'}},\varphi_* {\partial_{s}}_{\lvert_{x'}})}.\end{equation}
The argument of the square root is nothing but the metric $\tilde{g}$ induced  on $W$ by the conformal transformation $\varphi$,
\begin{equation}
\tilde{g}(U,V) = g(\varphi_* U, \varphi_* V),
\end{equation}
whose components are
$
\tilde{g}_{\mu\nu} = \partial_\mu {x'}^\alpha \partial_\nu {x'}^\beta
\eta_{\alpha\beta}$. Explicit computation yields
\begin{equation}
\label{metriqueinduite}
\tilde{g} = C^2 g
\end{equation}
with a conformal factor
\begin{equation}
\label{facteurconforme}
C(x) \doteq \frac{2L}{N(x)}.
\end{equation}
Hence (\ref{betaprim}) reads 
\begin{eqnarray}
\beta(x') &=& \sqrt{\tilde{g}_{\lvert_x}({\partial_{s}}_{\lvert_x},{\partial_{s}}_{\lvert_x})}\\
&=&C(x)\sqrt{g
  ({\partial_{s}}_{\lvert_x},{\partial_{s}}_{\lvert_x})}\\
&=& C(x)\norm{{\partial_{s}}_{\lvert_x}}
\end{eqnarray}
By \eqref{ddtdds}
and \eqref{unruhtemp},
$\partial_{s} = -\frac{2\pi}{a} \partial_{\tau}$
so that
\begin{equation}
\label{betaconform0}
\beta(x') = \frac{2\pi}{a} C(x).
\end{equation}
Hence the temperature at the point $x'$ on the orbit $H_D$ is the quotient of the Unruh temperature on the inverse image
$H_W=\varphi^{-1}(H_D)$ by the conformal factor at
$x = \varphi^{-1} (x').$
Note that, as expected, since $C(x)$ is not constant along an orbit $H_D$, neither is the temperature.

To check coherence with \cite{diamonds} one needs an expression of $\beta(x')$ as a function of the coordinates of $x'$. Let us note that any $x'\in D$ and $x\in
W$ are
uniquely determined by the values $a',a$ of the acceleration and 
$s, s'$ of the modular parameter.
 Using the commutativity
of the diagram
\begin{eqnarray}
W &\underset{\longleftarrow}{\varphi^{-1}}& D\\
\left. \sigma_W\right\uparrow& &\uparrow \sigma_D\\
W &\overset{\longleftarrow}{\varphi^{-1}}&D
\end{eqnarray}
obtained from (\ref{sigmad}), the inverse image of
$x' = (a',s')\in D$ is $x = (a,s)\in W$ with
\begin{eqnarray}
\label{eqs}
s &=& s',\\
\label{eqtau}
a &=& a'L + \sqrt{1+a'^2L^2}
\end{eqnarray}
where $a$ is obtained by inverting ~(\ref{eq:2}). Putting this value into (\ref{eq:6}) then (\ref{facteurconforme}), one obtains
\begin{equation}
 \label{betadiam}
 \beta(x') =  \beta(a',s') = \frac{2\pi L}{\sqrt{1 + a'^2 L^2} + \text{ch} \,2 \pi s'}
 \end{equation}
which is equation 53 of \cite{diamonds}. Initially this equation was obtained by computing $\beta$ directly from the $s'$ parametrization of $H_D$, without reference to the conformal map. We then observed as a striking fact that $\beta(a',s')$ was finite on all $D$, including the inertial trajectory.
We missed the following simple observation:
the conformal factor associated to a map from an unbounded region $W$ of Minkowski space-time to a bounded region $D$ cannot be infinite (roughly speaking, this is because the conformal factor $C$ has to ``contract'' $W$ to $D$). Hence $\beta$ which is 
proportional to $C$ has to be finite on all $D$ and $T_D = \beta^{-1}$ cannot vanish.

\section{Conclusion: conformal mapping of Unruh temperature}

A brief way to summarize this article is to rewrite
(\ref{betaconform0}) as
\begin{equation}
\label{main}
T_D (a',s') = \frac{T_U(a)}{C(a,s)}
\end{equation}
where $x'=(a',s')\in D$ and $x=(a,s) = \varphi^{-1}(x)\in W$. One may feel uneasy with the idea that $T_D$ is not constant along the trajectory of the observer:
what is the meaning of such an equilibrium state with non-constant temperature ? As noted in \cite{diamonds} $T_D(a',s')$ is not a constant along $H_D$ but 
it is remarkably flat on a
 plateau-region that increases with the size of the double-cone.
For most of its lifetime, the observer actually sees the vacuum as a
thermal state at temperature $T_D(a',0)$. This is only close to the extremities of the double-cone
that the temperature rapidly diverges.  
 In a sense by  the conformal map $\varphi$ the infinity of lifetime is mapped to a sharp divergence of the temperature at the extremities of the double-cone.

Another way to deal with this non-constant equilibrium temperature is to come back to the KMS condition. Since $\beta$ never vanishes, on a given $H_D$ it makes sense to slightly modify (\ref{ddtdds}) and define $\partial_t = -\beta(s')^{-1}\partial_{s'}$. Thus writing $\alpha_{-\beta(s) ds} \doteq \sigma_{ds}$  one obtains an equation similar to 
\eqref{kmstot} with $ds'$ instead of $s$. This infinitesimal KMS condition indicates that locally,
i.e. with respect to a small variation of time $dt = -\beta(s')ds'$, the vacuum 
has all the properties of a thermal state at temperature $\beta(s')^{-1}$. This idea has to be confronted to another definition of local equilibrium state in algebraic quantum field theory \cite{buchholzojima}. Work on that matter is in progress.

 From a physical point of view $T_D$ never vanishes 
because a non-eternal observer,  accelerated or not,
has access only to the degrees of freedom inside $D$ while there exist quantum
correlations 
between the inside and the outside of $D$. Thus the restriction of
the vacuum to the inside of $D$ is not a pure state. 
In the eternal case this analysis fits well with the
analogy with Hawking temperature and within appropriate coordinates
the edge of the wedge 
can really be seen as the horizon of a black hole.
For double-cones the situation is more problematic. Formally the
boundary of $D$ acts as a horizon, but there is no (at least to the
knowledge of the author) concrete realization of $D$ as the horizon of
a black hole. A theorem of Hawking indicates that the event horizon of
a stationary black hole must be a Killing-horizon, and the boundary of $D$ is not a Killing-horizon but
 only a conformal Killing-horizon. So a possible interpretation of $T_D$
 in terms of Hawking temperature would require non-stationary
 black-holes.
Note that in \cite{jacobkang} surface gravity, hence Hawking
temperature, is  defined for conformal Killing-horizons but only under
the requirement that the conformal factor be $1$ at infinity. However there is no proposition for an associated Unruh effect.
Here the situation is the opposite: the conformal factor vanishes at infinity (hence the divergence of
the temperature at the extremities of the double-cone) so that we cannot use \cite{jacobkang} to associate
some Hawking radiation to a double-cone. However we managed to define an Unruh temperature.

Another view on the problem is to observe that the conformal map
sends unextendible trajectories of observers in $W$ to extendible
orbits in $D$. So the interpretation of the edge of $D$ as a thermal
horizon for a finite-lifetime observer is delicate since one may
always imagine situations in which the observer is able to exchange information with a second observer 
whose line of universe goes through $D$. Also it should be noted that $T_D(a',s')$ cannot be seen
as
an instant temperature that the accelerated observer could measure at some proper-time
$\tau'\leq \tau'_0$. Doing so the observer would indeed be able to
deduce its lifetime from the inverse function $T_D^{-1}$, which is not acceptable from obvious causal reasons. Although this objection does 
not
survive a more rigorous analysis \cite{unruhcausal}, it  
still questions the physical relevance of double-cone-regions viewed as
 embedded regions within Minkowski space-time.

Alternatively one may consider $D$ as a universe by
itself, independently of any 
immersion in a larger space. Then the point is to study which of the states of $\aa(D)$
have a geometric modular action, then to single out by some intrinsic
properties the one corresponding
to the restriction of Minkowski vacuum. A characterization of states
that have a geometric modular action has been obtained for the wedge
algebra \cite{buchholzsummers} but the question is open
and difficult for the double-cone (see \cite{saffary,saffary2} for a
nice state-of-the-art discussion of this issue).

It could be also interesting to study whether Unruh's initial idea to compare different quantization schemes on $W$ 
might be adapted to double-cones.

Finally let us mention other double-cones whose modular group has a geometrical action, namely those associated to 
$2D$-boundary conformal field theory with boundary \cite{longo}. The point is to study \cite{diamod} the influence of the boundary on the temperature, with the idea that far from the boundary one should retrieve the temperature $T_D$ calculated in this paper.  

\section*{Acknowledgments}
\noindent Many thanks to K. H. Rehren and D. Buchholz for challenging
discussions, reading and corrections. Work partially supported by a EU Marie-Curie fellowship EIF-025947-QGNC.

\makeatletter
\renewcommand\@biblabel[1]{#1. }
\makeatother


\begin{thebibliography}{00}

\bibitem{bisognano1} J.~J.~ Bisognano, E.~H.~Wichman, {\it On the
duality condition for a hermitian scalar field}, J. Math. Phys. {\bf 16}
984 (1975).

 \bibitem{bisognano2} J.~J.~ Bisognano, E.~H.~Wichman,{\it On the duality condition for quantum fields},
Jour.~Math.~Phys.~{\bf 17} 303 (1976).

\bibitem{buchholzojima} D.~Buchholz, I. Ojima,
 {\it  Thermodynamic Properties of Non-equilibrium States in Quantum Field Theory},
Annals of Physics {\bf 297} 219-242 (2002). 

\bibitem{buchholz} D.~Buchholz, J. ~Schlemmer,
 {\it  Local Temperature in Curved Spacetime},
 Class.Quant.Grav.{\bf 24} F25-F31 (2007). 

\bibitem{buchholzsummers} D.~Buchholz, O. Dreyer, M. Florig, S. J. Summers,
{\it Geometric Modular Action and Spacetime Symmetry Groups},
Rev.Math.Phys. {\bf 12} (2000) 475-560.

\bibitem{connesmodul} A. Connes, \emph{Groupe modulaire d'une algèbre de von Neumann},
C. R. Acad. Sci. Paris, Sér. A-B, {\bf 274} (1972) A1923-A1926. 

\bibitem{carloconnes} A.~Connes, C.~Rovelli, {\it Von Neumann algebra
automorphisms and time-thermodynamics
relation in generally covariant quantum theories}, Class.~Quantum
Grav {\bf 11} 2899-2917 (1994).
 
\bibitem{davies} P.~C.~W.~Davies,
{\it Scalar Particle Production in Schwarzshild and Rindler
metrics}, J.Phys.{\bf A 8} 609 (1974).

\bibitem{dewitt} B. S. DeWitt, "Quantum gravity, the new
synthesis" in {\it General relativity; an Einstein centenary survey}
ed S. W. Hawking and W. Israel, Cambridge University Press (1979).

\bibitem{fulling} S.~A.~Fulling, {\it
Non-uniqueness of canonical field quantization in Riemannian
spacetime}, Phys. Rev.D {\bf 7} 2850 (1973).

 \bibitem{haag} R.~Haag, {\it Local Quantum Physics}, Springer (Berlin,
 1996).

\bibitem{hislop} P.~D.~Hislop, R.~Longo, {\it Modular structure
of the local algebras associated with free massless scalar field
theory}, Comm.~Math.~Phys.~{\bf 84} 71-85 (1982).

\bibitem{jacobkang} T. Jacobson, G. Kang, {\it Class. Quant. Grav.}{\bf 10}(1993) L201-L206.


\bibitem{longo} R. Longo, K.-H. Rehren, {\it Local Fields in Boundary Conformal QFT},
Reviews in Math. Phys. {\bf 16} (2004) 909-960.

\bibitem{diamod} R. Longo, P. Martinetti, K.-H. Rehren, in preparation.

\bibitem{louko} J. Louko, A. Satz, {\it How does often the Unruh-DeWitt
detector click ? Regularization by a spatial profile},
gr-qc/0606067.


\bibitem{unruhcausal} P. Martinetti, {\it A brief remark on Unruh effect and causality},
J. Phys.: Conf. Ser. {\bf 68} (2007) 012027.

\bibitem{thermalhorizon} P. Martinetti, {\it Is life a thermal horizon ?},
J. Phys.: Conf. Ser. {\bf 67} (2007) 01203.

\bibitem{diamonds} P. Martinetti, C. Rovelli, {\it Diamonds Temperature: Unruh effect
  for bounded trajectories and the thermal time hypothesis},
Class. Quant. Grav {\bf 20} (2003) 4919-4932, gr-qc/0212074. 


\bibitem{carlo93} C.~Rovelli, {\it Statistical mechanics of gravity
and the thermodynamical origin of time}, Class.~Quantum Grav {\bf
10} 1549-1566 (1993).

%

\bibitem{saffary} T. Saffary, {\it
 On the Generator of Massive Modular Groups},
 Lett.Math.Phys. {\bf 77} (2006) 235-248. 

\bibitem{saffary2} T. Saffary, {\it
 Modular Action on the Massive Algebra},
dissertation, math-ph/0512046 .

\bibitem{sewell1} G.~L.~Sewell, {\it Relativity of temperature and the Hawking effect},
Phys. Lett. {\bf 79 A} n. 1, 23 (1980).

\bibitem{sewell2} G.~L.~Sewell, {\it Quantum fields on
manifolds: PCT and gravitationally induced thermal states}, Annals
Physics {\bf 141}, 201-224 (1982).

\bibitem{tomtak} M.~Takesaki, {\it Tomita's Theory of Modular Hilbert
Algebras and its Applications},
Springer-Verlag (Berlin, 1970).

\bibitem{unruh} W.~G.~Unruh, {\it Notes on black hole evaporation},
Phys.~Rev.~D {\bf 14} 870 (1976). 



\end{thebibliography}
\end{document}